\documentclass[aps,prb,reprint,amsmath]{revtex4-1}
\usepackage{graphicx}
\usepackage{enumerate}
\usepackage{refcount}
\usepackage{fmtcount}
\usepackage{hyperref}
\usepackage{amssymb}
\usepackage{bbm}
\usepackage{bm}
\usepackage{refcount}
\usepackage{xcolor}

\hypersetup{pdfstartview=FitH,pdfpagemode=UseOutlines,colorlinks,
citecolor=blue,linkcolor=blue,urlcolor=blue}

\newcommand{\s}{\sigma}
\renewcommand{\t}{\theta}
\newcommand{\dg}{\dagger}
\newcommand{\sd}{\downarrow}
\newcommand{\su}{\uparrow}

\newcommand{\ket}[1]{\left|{#1}\right\rangle}
\newcommand{\bra}[1]{\left\langle{#1}\right|}

\begin{document}


\title{A measure of quantum correlations that lies approximately between entanglement and discord}

\author{Aaron Szasz}
	\email[]{aszasz@berkeley.edu}
	\affiliation{Department of Physics, University of California, Berkeley, California 94720, USA}
	\affiliation{Materials Sciences Division, Lawrence Berkeley National Laboratory, Berkeley, California 94720, USA}

\date{\today}

\begin{abstract}
When a quantum system is divided into two local subsystems, measurements on the two subsystems can exhibit correlations beyond those possible in a classical joint probability distribution; these are partially explained by entanglement, and more generally by a wider class of measures such as the quantum discord.  In this work, I introduce a simple thought experiment defining a new measure of quantum correlations, which I call the accord, and write the result as a minimax optimization over unitary matrices.  I find the exact result for pure states as a simple function of the Schmidt coefficients and provide a complete proof, and I likewise provide and prove the result for several classes of mixed states, notably including all states of two qubits and the experimentally relevant case of a pure state mixed with colorless noise.  I demonstrate that for two qubit states the accord provides a tight lower bound on the discord; for Bell diagonal states it is also an upper bound on entanglement.  
\end{abstract}

\maketitle


\section{Introduction\label{section:introduction}}

The classic example of an entangled quantum state is the singlet state of two spin-1/2 systems,
\begin{equation}
\ket{\Psi^-} = \frac{1}{\sqrt 2}\big(\ket{\su\sd}-\ket{\sd\su}\big),\label{eq:singlet}
\end{equation}
where $\{\ket{\su},\ket{\sd}\}$ is an orthonormal basis for the Hilbert space for each spin.  This is a maximally entangled state, meaning that measurements of the two spins, when made along the same spatial axis, will always be perfectly correlated, even if the spins are space-like separated when the measurements occur.  The opposite case is a product state, in which the two parts of the system can be described completely independently.  Partially entangled states lie between these two extremes, and substantial effort has gone into finding ways of quantifying the precise degree of entanglement and correlation in such states.\cite{Plenio2007,Horodecki2009,Modi2012}

One view is that entanglement is a form of nonlocality.  If this were true, an entangled state would violate some Bell-type inequality\cite{Bell1964} that is satisfied by any local hidden variable model (LHVM), such as the Clauser-Horne-Shimony-Holt (CHSH) inequality\cite{CHSH1969} for a system of two spin-1/2 subsystems, or similar inequalities involving more\cite{Mermin1990,Ardehali1992,Son2006} or higher-dimensional\cite{Collins2002,Masanes2003,Son2006} subspaces.  States can be classified by whether or not they violate such an inequality, which all non-product pure states do\cite{Gisin1991}.  The degree of nonlocality can also be quantified, for example by the maximal amount of random noise that can be added to the state such that it still cannot be described by a LHVM\cite{Kaszlikowski2000}.

Alternatively, the entanglement of a state can be quantified by the number of singlet states, of the form \eqref{eq:singlet}, to which it is equivalent.\cite{Bennett1996a,Plenio2007}  For example, one can ask how many singlets, $m$, can be made from $n$ copies of the given state in the limit that $n$ becomes large; the ratio $m/n$ is called the distillable entanglement.\cite{Bennett1996}  Such measures of singlet equivalence are equivalent to the entanglement entropy\cite{Bennett1996a,Popescu1997} on pure states and satisfy certain axioms;\cite{Vedral1997,Vidal2000,Donald2002} these are formally known as entanglement measures.  For mixed states there are many inequivalent measures such as entanglement of formation\cite{Bennett1996,Hayden2001}, the aforementioned distillable entanglement, entanglement of purification\cite{Terhal2002}, and logarithmic negativity\cite{Vidal2002,Plenio2005} that give different orderings on the set of states\cite{Eisert1999,Eltschka2015}.

Quantum correlations can also be understood through their ability to act as a resource for tasks in quantum computation.  One prominent example is quantum teleportation\cite{Bennett1993}, in which an entangled state shared between two subsystems can be used to transfer the state of a particle from one subsystem to the other.  The average fidelity for such a transfer is linearly related to the singlet fraction of the shared entangled state, which is its largest overlap with a maximally entangled state with the same subspace dimensions\cite{Horodecki1996b}.

Among pure states, entangled states as identified via the entanglement entropy are precisely the same as those that violate Bell-type inequalities\cite{Gisin1991,Popescu1992} and as those that allow teleportation with a greater fidelity than is possible by any classical strategy\cite{Horodecki1996b,Banaszek2000}.  

For mixed states, this is no longer the case.  There are entangled states that admit a LHVM and do not violate the CHSH inequality\cite{Werner1989} even with a sequence of measurements\cite{Popescu1995} and similarly there are states that admit a LHVM but can still be used for quantum teleportation with greater fidelity than is possible by any classical strategy\cite{Popescu1994}.  At the same time, there are computational tasks with quantum advantages over classical algorithms that cannot be explained by entanglement\cite{Knill1998,Lanyon2008,Bromley2017}, so a different notion of quantumness versus classicality is needed.

The quantum discord, introduced independently by Henderson and Vedral\cite{Henderson2001} and Ollivier and Zurek\cite{Ollivier2002}, quantifies the notion of nonclassicality in mixed states; given a state shared between two subsystems, the discord computes how much the state of one subsystem is necessarily modified, on average, by a measurement on the other.  The discord and its variants, including geometric discord\cite{Dakic2010,Luo2010}, diagonal discord\cite{Lloyd2015,Liu2017}, and others\cite{Luo2008a,Modi2010,Modi2012,Farace2014,Spehner2014,Adesso2016}, are nonzero on most separable states\cite{Ferraro2010}.  There is strong evidence to suggest that discord is the relevant resource for a variety of quantum computational tasks.\cite{Datta2008,Lanyon2008,Cavalcanti2011,Passante2011,Dakic2012,Gu2012,Adesso2016,Braun2018}  

In this paper, I present a new measure of quantum correlations, the accord, defined by a simple thought experiment.  The rough idea is that entanglement between two subsystems means that there is an inescapable correlation between measurements made on the two; imagining a game in which the holder of one subsystem, Bob, tries to make his measurements as unpredictable as possible to the holder of the other, Alice, the measure is the (rescaled) probability that Alice is able to guess Bob's measurements correctly, despite Bob's best efforts to prevent this.  

The primary advantage of the accord compared with existing measures of quantum correlations is that, because it is defined directly in terms of a simple experimental procedure, it has a clear intuitive meaning; at the same time, as I demonstrate below, the accord can be efficiently computed for wide classes of states.  The existence of such a measure, namely one that is both intuitively understandable and efficient to compute, is important because it allows for new insights into other, more commonly used measures such as entanglement and discord.  In particular, I both show that the accord provides a lower bound on discord and demonstrate the remarkable and surprising fact that there are states with zero accord and nonzero entanglement, thus revealing a new type of ``hidden entanglement.''

The organization of the paper is as follows: in section \ref{section:thought_experiment}, I motivate the thought experiment and use it to formally define the accord as a variational optimization over unitary matrices.  In section \ref{section:pure_states} I evaluate the accord for pure states and prove the result, and in section \ref{section:mixed_states} I prove some results for mixed states, including a simple and efficient prescription to compute the measure on all two qubit states.  In section \ref{section:comparison} I compare the accord with existing measures from the literature.  Finally, in section \ref{section:discussion} I conclude with a summary and a discussion of the significance of the results.


\section{The thought experiment\label{section:thought_experiment}}

I begin with an example for motivation.  Two observers, Alice ($A$) and Bob ($B$), each hold one qubit, realized as a spin-1/2 system, and the two qubits are in some possibly entangled state.  Consider in particular the following two pure states:
\begin{align}
\ket{\Phi^+} & = \frac{1}{\sqrt 2}\big(\ket{\su\su}+\ket{\sd\sd}\big),\label{eq:bell_phi_p}\\
\ket{\psi_\text{sep}} & = \ket{\su\su}\label{eq:unent},
\end{align}
where $\ket\su$ and $\ket\sd$ are the eigenstates of the operator $S_z$.
The first state is maximally entangled and the second is separable, so measurements made by $A$ and $B$ should be more correlated in the first state; however, if $A$ and $B$ both naively measure $S_z$, their measurements will be perfectly correlated in either case.  Likewise, if $B$ chooses to measure $S_x$ while $A$ still chooses to measure $S_z$, the measurements in both cases will be completely uncorrelated.  

But now suppose that $A$ knows the initial state and also knows $B$'s measurement axis.  In that case, if $B$ chooses to measure $S_x$, when the shared state is $\ket{\Phi^+}$ $A$ can choose to also measure $S_x$, in which case their measurements again become perfectly correlated, but when it is $\ket{\psi_\text{sep}}$, their measurements will be completely uncorrelated no matter what axis $A$ chooses for her measurement.

In other words, the state $\ket{\Phi^+}$ can be said to be maximally entangled because no matter what spin component $B$ chooses to measure, $A$ can always choose one to achieve perfect correlation between their measurements, while the state $\ket{\psi_\text{sep}}$ is separable because $B$ can choose a spin component for which, no matter what component $A$ chooses, their measurements will be completely uncorrelated.  For a partially entangled state between these two extremes, the degree of entanglement is characterized by how correlated $A$ can force their measurements to be by an appropriate choice of measurement axis, even in the worst case of the choice made by $B$.

\subsection{Formal statement, version 1\label{section:formal_1}}
I now formalize the above intuition.  The setup is as follows: two observers, Alice ($A$) and Bob ($B$), share many copies of a quantum state, $\rho$, in the Hilbert space $\mathcal{H}=\mathcal{H}_A\otimes\mathcal{H}_B$.  The subspaces held by the two observers, $\mathcal{H}_A$ and $\mathcal{H}_B$, are both $d$-dimensional; let $\{\ket 0_A,\cdots,\ket{d-1}_A\}$ be an orthonormal basis for $\mathcal{H}_A$ and $\{\ket 0_B,\cdots,\ket{d-1}_B\}$ for $\mathcal{H}_B$.  $A$ and $B$ are each capable of applying any unitary transformation $U\in U(d)$ to their respective subspace, and each has a device for perfect projective measurements of some operator that is diagonal in the specified basis states and nondegenerate, for example $\hat{n}$ defined by $\hat{n}\ket n = n\ket n$.  An application of some $U$ before measurement can be thought of as making the measurement in a different basis (eg $S_x$ vs $S_z$ in the example above).

In the example, $A$ was able to pick the right measurement basis to guarantee correlations in the maximally entangled state only because she knew both (1) the initial state and (2) $B$'s choice of basis.  Likewise, $B$ was only able to pick a basis to guarantee a lack of correlation for the separable state because (3) he knew the initial state.  For this first formulation of the thought experiment I assume (1)-(3); these assumptions are dangerously strong, but I will show in the second formulation that they are not actually necessary.

I now define the correlation measure by a procedure which for clarity I present in reverse chronological order:
\begin{enumerate}
\item[3.] For fixed $\rho$, $U_A$, and $U_B$, $A$ measures $\hat{n}$ after applying $U_A$ and $B$ measures $\hat{n}$ after applying $U_B$.  The measurement coincidence probability, or MCP, is the probability that the two measurements agree.
\item[2.] Prior to step 3, $A$ chooses $U_A$ to maximize the MCP, given her knowledge of (assumption 1) $\rho$ and (assumption 2) $U_B$.
\item[1.] Prior to step 2, $B$ chooses $U_B$ to minimize $A$'s maximized MCP, given his knowledge (assumption 3) of $\rho$.  He communicates this choice to $A$ for use in step 2.
\end{enumerate}

\noindent The value of the MCP, given $U_A$, $U_B$, and $\rho$, is
\begin{equation}
\sum_{n_A} P(\hat{n}_A=n_A)\times P(\hat{n}_B=n_A|\hat{n}_A=n_A) \label{eq:prob_eq_1}
\end{equation}
where
\begin{widetext}
\begin{equation}
P(\hat{n}_B=n_A|\hat{n}_A=n_A) = P(\hat{n}_A=n_A 
,\hat{n}_B=n_A)/P(\hat{n}_A=n_A)
\end{equation}
and
\begin{equation}
P(\hat{n}_A=n_A 
,\hat{n}_B=n_A) = \text{Tr}\left(\,\ket{n_A,n_A}\bra{n_A,n_A}\,(U_A\otimes U_B)\,\rho\,(U_A\otimes U_B)^\dg\right)
\end{equation}
where $\ket{n_A,n_A}$ is shorthand for $\ket{n_A}_A\otimes\ket{n_A}_B$.  Thus the optimized MCP, or OMCP, is
\begin{equation}
\text{OMCP} \equiv \min_{U_B}\!\left(\!\max_{U_A}\!\left(\sum_{n=0}^{d-1}\bra{n,n}(U_A\otimes U_B)\,\rho\,(U_A\otimes U_B)^\dg\ket{n,n}\right)\!\!\right).\label{eq:def_OMCP}
\end{equation}
\end{widetext}

As I show in section \ref{section:lower_bound} below, $1/d\leq \text{OMCP} \leq 1$, so to compare with other measures it will be useful to also define a rescaled version that runs from 0 to 1 for any $d$,
\begin{equation}
\text{Accord} \equiv \frac{d}{d-1}\left(\text{OMCP}-\frac{1}{d}\right).\label{eq:def_accord}
\end{equation}  
The name is of course a reference both to the similarity to the discord and to the fact that the measure is based on agreement between measurements.

\subsection{Formal statement, version 2\label{section:formal_2}}

The first statement of the thought experiment can be viewed as a game: the first player, $A$, tries to maximize her score by making the the two parties' measurements agree, while the second player, $B$, tries to minimize $A$'s score by making the measurements uncorrelated.  This formulation requires the assumptions (1)-(3) so that both players can make optimal choices of their measurement bases.

The assumptions can be relaxed by viewing the optimization over unitary matrices in equation \eqref{eq:def_OMCP} not as an explicit choice of the optimal change of basis, but rather as an optimization of the observed measurement coincidence probability over a large set of randomly chosen (or otherwise uniformly distributed) unitaries.  The correlation measure can thus be defined according to the following procedure:
\begin{enumerate}
\item $B$ selects some random set of $N_B$ unitary transformations.
\item For each $U_B$ selected by $B$, $A$ selects $N_A$ random unitary transformations.
\item For each pair $(U_B,U_A)$, $A$ and $B$ apply their respective transformations to many copies of the state $\rho$ and measure $\hat{n}$, then record the fraction of the time that the two measurements agree.
\item For each $U_B$, they take the maximum over all the coincidence probabilities from step 3 with that $U_B$.
\item Finally, they take the minimum value from step 4 over all choices of $U_B$.
\end{enumerate}
This procedure evidently leads, in the limit that $N_A$ and $N_B$ become large, to the exact same final expression given in equation \eqref{eq:def_OMCP}, and as promised assumptions (1)-(3) are no longer needed.  In principle this formulation allows for a direct experimental probe of entanglement in an unknown state, requiring only the ability to apply random one-subsystem unitaries and to prepare many copies of the desired state, but the number of measurements required is probably too large to be practical compared with a full state tomography\cite{James2001}.

\subsection{Extension to unequal subspace dimensions\label{section:unequal_dims}}

The MCP is defined in terms of the probability that the measurements made by $A$ and $B$ agree, which requires that they be able to make equivalent measurements, ie. that the two subspaces should be isomorphic.  It is thus not obvious how to extend the measure to the case of unequal subspace dimensions.

Supposing that the two dimensions are $d_1 > d_2$, one option would be to arbitrarily select $d_2$ of the $d_1$ states as the ones that should match; the result will not depend on which ones are chosen, since whichever party has the subspace of dimension $d_1$ can apply a unitary to permute their basis states.

To formalize this, one can use equation \eqref{eq:def_OMCP} for a $d_1^2\times d_1^2$ density matrix, with $d_1(d_1-d_2)$ rows and columns equal to 0, and with the unitary matrices for whichever party has the smaller subspace restricted to act as the identity on the corresponding $d_1-d_2$ dimensions (thus preserving the zero rows and columns in $\rho$).

In this paper I will not consider this case further.


\section{Pure states\label{section:pure_states}}

In the special case that the state $\rho$ is in fact a pure state, $\rho=\ket\psi\bra\psi$, equation \eqref{eq:def_OMCP} can be evaluated explicitly, as I now demonstrate.

The first step is to make use of the Schmidt decomposition: given any pure state $\ket\psi$ in $\mathcal{H}_A\otimes\mathcal{H}_B$, there exist unitary matrices $\tilde{U}_A$ and $\tilde{U}_B$ and nonnegative numbers $\{c_0,\cdots,c_{d-1}\}$ satisfying $\sum c_i^2 = 1$, such that
\begin{equation}
\ket\psi = \big(\tilde{U}_A\otimes \tilde{U}_B \big)\sum_{i=0}^{d-1} c_i \ket{i}_A\otimes\ket{i}_B.\label{eq:def_Schmidt_decomp}
\end{equation}
The coefficients $c_i$ are unique given $\ket\psi$, although $\tilde{U}_A$ and $\tilde{U}_B$ are not.

Using equation \eqref{eq:def_OMCP} with $\rho=\ket\psi\bra\psi$ with $\ket\psi$ in this Schmidt-decomposed form, $\tilde{U}_A$ and $\tilde{U}_B$ only appear in the combinations $U_A\tilde{U}_A$ and $U_B\tilde{U}_B$; since the OMCP involves optimization over both $U_A$ and $U_B$, $\tilde{U}_A$ and $\tilde{U}_B$ may each be assumed without loss of generality to be the $d\times d$ identity matrix.  In other words, the OMCP depends only on the Schmidt coefficients $\{c_i\}$, and thus the state $\ket\psi$ can be assumed without loss of generality to be of the form
\begin{equation}
\ket\psi = \sum_{i=0}^{d-1} c_i\ket{i}_A\ket{i}_B.\label{eq:Schmidt_decomp_psi}
\end{equation}
Equation \eqref{eq:def_OMCP}, when evaluated for $\rho=\ket\psi\bra\psi$ with $\ket\psi$ from equation \eqref{eq:Schmidt_decomp_psi}, gives
\begin{equation}
\text{OMCP} = \frac{1}{d}\left(\sum_{i=0}^{d-1} c_i\right)^2.\label{eq:OMCP_pure}
\end{equation}
In the proceeding sections, I provide an intuitive picture to explain this result, followed by a complete proof.

\subsection{Intuitive picture\label{section:pure_intuition}}

To build intuition, I begin with the case of $d=2$.  Consider the state
\begin{equation}
\ket\psi = c_0\ket{00} + c_1\ket{11};
\end{equation}
if $A$ and $B$ each measure immediately without applying a unitary first, their measurements will be in perfect agreement.  Thus it is intuitively reasonable that to reduce this coincidence probability, $B$'s goal in the first formulation of the thought experiment, he ought to try to get as far from this basis as possible.  Viewing the qubits as spin-1/2 systems with the state originally specified in the $S_z$ basis, $B$'s optimal measurement axis would be one in the $xy$-plane.

I provide two examples: if $B$ chooses to measure along $x$ or along $y$, that is equivalent to applying the unitary matrix
\begin{equation}
U_B^{(x)}=\frac{1}{\sqrt 2}\left(\begin{array}{cc}
1 & 1\\ 1& -1
\end{array}\right)\,\,\,\text{or}\,\,\,\,
U_B^{(y)}=\frac{1}{\sqrt 2}\left(\begin{array}{cc}
1 & 1\\ i& -i
\end{array}\right),
\end{equation}
respectively.  With these choices, if $A$ naively chooses to measure in the $S_z$ basis, the measurement coincidence probability will be only 50\%.  However, if $A$ instead chooses to use optimal bases, namely
\begin{equation}
U_A^{(x)}=\frac{1}{\sqrt 2}\left(\begin{array}{cc}
1 & 1\\ 1& -1
\end{array}\right),\,\,\,\,\,
U_A^{(y)}=\frac{1}{\sqrt 2}\left(\begin{array}{cc}
1 & 1\\ -i& i
\end{array}\right),
\end{equation}
then the state $\ket\psi$ becomes in the two cases 
\begin{align}
\ket\psi_{xx} & = \frac{c_0+c_1}{2}\left(\ket{00}+\ket{11}\right) + \frac{c_0-c_1}{2}\left(\ket{01}+\ket{10}\right)\\
\ket\psi_{yy} & = \frac{c_0+c_1}{2}\left(\ket{00}+\ket{11}\right) + i\frac{c_0-c_1}{2}\left(\ket{01}-\ket{10}\right).
\end{align}
Either way, the probability that $A$ and $B$'s measurements will be the same is exactly $(c_0+c_1)^2/2$.  That the specified $U_A$ and $U_B$ are optimal is by no means obvious but can be demonstrated by writing fully general unitaries and explicitly performing the optimization.

For $d>2$, some lessons should carry over: (1) $B$'s measurement basis should maximally mix his original basis states, and (2) an optimal choice for $A$ is $U_A=U_B^\ast$.  (The second point turns out not to be true for general $U_B$, but it is true when $B$ makes an optimal choice.)  With this in mind, we consider the state
\begin{equation}
\ket\psi = c_0\ket{00} + \cdots c_{d-1}\ket{d-1,d-1}.
\end{equation}
$B$ maximally mixes before measuring by applying the change of basis unitary with elements
\begin{equation}
\left[U_B\right]_{jk} = \omega_d^{jk}/\sqrt{d}\label{eq:optimal_UB}
\end{equation}
where $\omega_d$ is the $d$th root of unity $\exp(2\pi i/d)$ and $j$ and $k$ run from 0 to $d-1$, while $A$ tries to unmix using $U_A = U_B^\ast$.  The resulting state is
\begin{align}
\ket\psi & = \frac{1}{d}\sum_{j}c_j\left(\sum_k e^{-2\pi jk/d}\ket{k}\right)\left(\sum_m e^{2\pi jm/d}\ket{m}\right)\\
& = \frac{1}{d}\sum_{km}\left(\sum_j c_j e^{2\pi j(m-k)/d}\right)\ket{km}.
\end{align}
The largest coefficients are those with no destructive interference, $m-k=0$, and these are precisely the ones we wanted to maximize, corresponding to agreement between $A$ and $B$'s measurements; those coefficients are all equal, with a value of $(\sum_j c_j)/d$.  The overall probability that the two measurements are equal is the sum of the squares of these coefficients, precisely matching equation \eqref{eq:OMCP_pure}.

\subsection{Proof\label{section:pure_proof}}

I now prove the result.  To do so, I rewrite the OMCP for pure states in two equivalent forms:
\begin{align}
\text{OMCP} & = \min_{U_B}\left(\max_{U_A}\left(|\!|
(U_A\circ U_B)\mathbf{c}|\!|^2\right)\right)\label{eq:OMCP_simple_1}\\
 & = \min_{U_B}\left(\max_{U_A}\left(\text{Tr}((U_A\Lambda U_B^T)\circ(U_A^\ast\Lambda U_B^\dg))\right)\right).\label{eq:OMCP_simple_2}
\end{align}
Here $\circ$ is the elementwise, or Hadamard, product, $\mathbf{c}$ is a vector whose entries are the Schmidt coefficients $\{c_i\}$, and $\Lambda$ is a diagonal matrix whose diagonal entries are again the Schmidt coefficients.  That these are equivalent to the OMCP is proven in Appendix \ref{appendix:OMCP_simple_proof}.  Using these expressions, I prove the result in two steps.

\vspace{\baselineskip}

\noindent{\bf Step 1:} $\text{OMCP} \geq (\sum c_i)^2/d$

\smallskip

For any fixed $U_B$, 
\begin{equation}
\max_{U_A}\left(|\!|(U_A\circ U_B)\mathbf{c}|\!|^2\right) \geq |\!|
(U_B^\ast\circ U_B)\mathbf{c}|\!|^2
\end{equation}
since $U_B^\ast$ is included as a possible $U_A$ on the left-hand side, and thus 
\begin{equation}
\min_{U_B}\left(\max_{U_A}\left(|\!|
(U_A\circ U_B)\mathbf{c}|\!|^2\right)\right) \geq \min_{U_B}\left(|\!|
(U_B^\ast\circ U_B)\mathbf{c}|\!|^2\right).
\end{equation}
It therefore suffices to show that  
\begin{equation}
|\!|
(U_B^\ast\circ U_B)\mathbf{c}|\!|^2 \geq \frac{1}{d}\left(\sum_{i=0}^{d-1} c_i\right)^2\label{eq:step1_suffices}
\end{equation}
for all $U_B$.

To do so, I use the lemma\cite{Taskara2013}
\begin{widetext}
\begin{equation}
\text{Tr}(A)\text{Tr(B)}=d\,\text{Tr}(A\circ B) - \sum_{i=1}^{d-1}\sum_{j=i+1}^d(a_{ii}-a_{jj})(b_{ii}-b_{jj})\label{eq:Had_trace_lemma}
\end{equation}
(in Appendix \ref{appendix:Had_tr_pf} I present an alternate proof to the one in reference \onlinecite{Taskara2013}) with equation \eqref{eq:OMCP_simple_2}, finding that for any unitary $U$
\begin{align}
|\!|
(U^\ast\!\circ U)\mathbf{c}|\!|^2 = \text{Tr}\!\left((U^\ast\Lambda U^T)\circ(U\Lambda U^\dg)\right)
& = \frac{1}{d}\!\left[\text{Tr}\!\left(U^\ast\Lambda U^T\right)\text{Tr}\!\left(U\Lambda U^\dg\right) + \sum_{i=1}^{d-1}\sum_{j=i+1}^d\!\big|(U\Lambda U^\dg)_{ii}-(U\Lambda U^\dg)_{jj}\big|^2\right]\!\geq\! \frac{\text{Tr}\!\left(\Lambda\right)^2}{d}.
\end{align}
\end{widetext}
The trace of $\Lambda$ is just the sum of the Schmidt coefficients, thus proving equation \eqref{eq:step1_suffices}; note that as demonstrated in the previous section the bound in equation \eqref{eq:step1_suffices} is achieved by the matrix $U_B$ given in equation \eqref{eq:optimal_UB} above.

\hfill $\blacksquare$ (Step 1)

\vspace{\baselineskip}

\noindent{\bf Step 2:} $\text{OMCP} \leq (\sum c_i)^2/d$

\smallskip

For any fixed unitary matrix $U_B^0$,
\begin{equation}
\min_{U_B}\left(\max_{U_A}\left(|\!|
(U_A\circ U_B)\mathbf{c}|\!|^2\right)\right) \leq \max_{U_A}\left(|\!|
(U_A\circ U_B^0)\mathbf{c}|\!|^2\right),
\end{equation}
so it suffices to find some $U_B^0$ such that
\begin{equation}
|\!|(U_A\circ U_B^0)\mathbf{c}|\!|^2 \leq \frac{1}{d}\left(\sum_{i=0}^{d-1} c_i\right)^2\label{eq:step_2_suffices}
\end{equation}
for all $U_A$.  Unsurprisingly, this is again achieved by the $U_B$ given in equation \eqref{eq:optimal_UB}, as I show now.  With that choice of $U_B^0$, we get
\begin{equation}
|\!|(U\circ U_B^0)\mathbf{c}|\!|^2 = \frac{1}{d}\sum_{jk}c_j c_k \left[\sum_l e^{i(2\pi/d)l(k-j)} U_{lj}^\ast U_{lk}\right].
\end{equation}
The expression in square brackets can be written as an inner product between two vectors:
\begin{align}
\sum_l e^{i(2\pi/d)l(k-j)} U_{lj}^\ast U_{lk} & = \langle v|w\rangle,\\
v_l & = e^{i(2\pi/d)lj}U_{lj},\\
w_l & = e^{i(2\pi/d)lk}U_{lk}.
\end{align}
Using the Cauchy-Schwarz inequality, this inner product satisfies $|\langle v|w\rangle| \leq |\!|v|\!|\,|\!|w|\!|$, and since $U$ is unitary, each row and column of $U$ is a normalized vector so that $|\!|v|\!|=|\!|w|\!|=1$.  Thus
\begin{equation}
|\!|(U\circ U_B)\mathbf{c}|\!|^2 \leq \frac{1}{d}\sum_{jk}c_j c_k = \frac{1}{d}\left(\sum_{i=0}^{d-1} c_i\right)^2,
\end{equation}
\begin{minipage}{\columnwidth}
completing the proof of equation \eqref{eq:step_2_suffices}.  The bound in that equation is achieved by $U_A={U_B^0}^\ast$.

\hfill $\blacksquare$ (Step 2)
\end{minipage}

\vspace{\baselineskip}

\noindent In combination, the two inequalities $\text{OMCP} \geq (\sum c_i)^2/d$ and $\text{OMCP} \leq (\sum c_i)^2/d$ prove equation \eqref{eq:OMCP_pure}.  


\section{Mixed states\label{section:mixed_states}}

I now turn to the more general case of mixed states.  The problem of calculating entanglement measures on mixed states is notoriously difficult, with only a few, such as negativity, being computationally tractable in general.\cite{Huang2014}  The OMCP (equivalently, the accord), too, is quite difficult to evaluate for general mixed states, and I do not have a general solution analogous to equation \eqref{eq:OMCP_pure} for pure states.  However, substantial analytical progress is still possible.  In particular, I prove universal upper and lower bounds on the OMCP, and I present exact results for several particularly important classes of mixed states, most notably all states of two qubits and all mixtures of a pure state with colorless noise.

\subsection{Upper and lower bounds\label{section:lower_bound}}
The upper bound of the OMCP is exactly 1, since it is defined as an optimized probability. This is achieved by maximally entangled pure states as demonstrated in the previous section. 

The lower bound is $1/d$; this is the probability that $A$ and $B$'s measurements agree when their shared state $\rho$ is completely uncorrelated.  The presence of correlations should not decrease this probability if $A$ makes a good choice of basis, so it is natural to conjecture that for any $\rho$, $\text{OMCP}\geq 1/d$.  Furthermore, a measurement coincidence probability of at least $1/d$ should be achievable by a strategy that does not depend at all on any correlations between the two subsystems that might happen to exist.  This is indeed the case.

\smallskip

\noindent {\bf Proof:} Let $U_B$ be fixed.  Then the probability that $B$ measures outcome $i$ is $P(\hat{n}_B=i) = \bra{i}U_B\rho_B U_B^\dg\ket{i}$ where $\rho_B$ is the reduced density matrix on $B$ given by the partial trace of $\rho$ over subsystem $A$.  Assume without loss of generality that $P(\hat{n}_B=0)\geq P(1)\geq\cdots \geq P(d-1)$.

Likewise, $P(\hat{n}_A=i) = \bra{i}U_A\rho_A U_A^\dg\ket{i}$.  When $U_A$ is the identity matrix, there is some ordering of the probabilities, $P(\hat{n}_A=\s_0)\geq P(\s_1) \geq \cdots \geq P(\s_{d-1})$, where $\s$ is some permutation.  $A$ can then choose $U_A$ to be the permutation matrix for $\s^{-1}$, in which case $P(\hat{n}_A=0)\geq P(1)\geq\cdots \geq P(d-1)$.

For that choice of $U_A$, the probability that $A$ and $B$'s measurements agree is 
\begin{equation}
\sum_{i=0}^{d-1} P(\hat{n}_A=i)\times P(\hat{n}_B=i)\geq \sum_{i=0}^{d-1} P(\hat{n}_A=i)\times\frac{1}{d}
\end{equation}
where the inequality follows by viewing each side as a weighted sum of the probabilities on subsystem $A$; going from the left-hand side to the right-hand side increases the weights given to the smaller probabilities and decreases the weights given to the greater ones.  The factor of $1/d$ can then be pulled out of the sum, and the sum on probabilities for subsystem $A$ of course gives 1.

This shows that for any $\rho$ and any $U_B$, there exists some $U_A$ (in fact, some permutation matrix) such that $\text{MCP}\geq 1/d$.  Thus the maximum over all $U_A$ is also at least this large, and since this is true for all $U_B$ the minimum over $U_B$ is as well.  In other words, for any state $\rho$, $\text{OMCP}\geq 1/d$. 

\hfill$\blacksquare$

\smallskip

Thus $1/d\leq\text{OMCP}\leq 1$, and hence the accord of equation \eqref{eq:def_accord} runs from 0 to 1 as intended.

\subsection{Classical states}

Classical states are those for which there exists a complete set of projective measurements that leave the state invariant; these are precisely the states of the form\cite{Luo2008a,Modi2010,Spehner2014,Adesso2016}
\begin{equation}
\rho = \sum_{i,j=0}^{d-1} a_{ij} \ket{\psi_i^A}\bra{\psi_i^A}\otimes\ket{\psi_j^B}\bra{\psi_j^B}
\end{equation}
where $\left\{\ket{\psi_i^A}\right\}$ is some orthonormal basis of $\mathcal{H}_A$ and likewise for $B$. That is, $\rho$ is diagonal in a basis of orthogonal separable states.  In this case, the OMCP is exactly $1/d$, corresponding to random chance and a total lack of correlation.  

\smallskip

\noindent{\bf Proof:} Substituting $\rho$ into the MCP from equation \eqref{eq:def_OMCP}, the MCP factorizes for each term in the sum on $i$ and $j$:
\begin{equation}
\text{MCP} = \sum_{i,j=0}^{d-1}a_{ij}\sum_{n}\big|\bra {n} U_A\ket{\psi_i^A}\big|^2\times \big|\bra{n} U_B\ket{\psi_j^B}\big|^2.
\end{equation}

Because the states $\ket{\psi_i^A}$ form an orthonormal basis for $\mathcal{H}_A$, there exists a change of basis matrix $\tilde{U}_A$ such that $\ket{\psi_i^A}=\tilde{U}_A\ket{i}_A$ for all $i$, where $\ket{i}_A$ is an element of the same standard basis as $\ket{n}_A$.  Since $U_A$ is optimized over, $U_A\tilde{U}_A$ can be renamed to $U_A$; in other words, we can assume without loss of generality that $\ket{\psi_i^A}=\ket{i}_A$, in which case the expression $\bra{n} U_A\ket{\psi_i^A}$ is nothing but the matrix element $U^A_{ni}$.  Making this substitution and following the same steps for $B$, the OMCP becomes
\begin{equation}
\min_{U_B}\left(\max_{U_A}\left(\sum_{i,j} a_{ij}\sum_n \big|U^A_{ni}\big|^2\times \big|U^B_{nj}\big|^2\right)\right).
\end{equation} 
Now suppose that $U_B$ is any unitary for which all elements are equal in magnitude, $\big|U^B_{nj}\big|^2 = 1/d$.  Then the expression to optimize is just
\begin{equation}
\frac{1}{d}\sum_{i,j} a_{ij}\sum_n \big|U^A_{ni}\big|^2,\label{eq:sep_ortho_step}
\end{equation}
and the inner sum is exactly the norm of the $n$th row of $U_A$, which is 1.  This leaves $(\sum a_{ij})/d$, and because $\rho$ is a normalized density matrix the sum on $a$ is 1 as well.  In other words, there exists a $U_B$ for which
\begin{equation}
\max_{U_A}\left(\sum_{i,j} a_{ij}\sum_n \big|U^A_{ni}\big|^2\times \big|U^B_{ni}\big|^2\right) = \frac{1}{d}
\end{equation}
and thus the minimum over $U_B$ is no larger than this.  So $\text{OMCP}\leq 1/d$. 

The opposite inequality, that $\text{OMCP}\geq 1/d$, has already been shown above to hold for any state $\rho$.  The two inequalities, taken together, prove that $\text{OMCP}=1/d$ as claimed.

\hfill $\blacksquare$

\smallskip

To summarize, the OMCP achieves its minimum possible value, $1/d$, when $\rho$ is diagonal in a basis of orthogonal separable states.  One might think that this implies that the OMCP achieves its minimum value on all separable states, but that is not the case, as I show below.

\subsection{Pure states with colorless noise}

One particularly experimentally relevant class of mixed states consists of pure states mixed with colorless noise; the latter is represented by the maximally mixed state, $\mathbbm{1}/d^2$, where $\mathbbm{1}$ is the $d^2\times d^2$ identity matrix.  Such a state is written as
\begin{equation}
\rho = x\ket{\psi}\bra{\psi} + (1-x)\frac{\mathbbm{1}}{d^2}\label{eq:pure_plus_noise_rho}
\end{equation}
for some $x\in[0,1]$.  The evaluation of the OMCP is actually quite easy: since the second term is invariant under conjugation by any unitary transformation, the OMCP becomes
\begin{widetext}
\begin{equation}
x\times\min_{U_B}\!\left(\!\max_{U_A}\!\left(\!\sum_{n}\bra{n,n}(U_A\otimes U_B)\,\ket\psi\bra\psi\,(U_A\otimes U_B)^\dg\ket{n,n}\right)\!\!\right) + \frac{(1-x)}{d}=x\times\text{OMCP}\left(\ket\psi\right)+\frac{(1-x)}{d}.\label{eq:pure_plus_noise_OMCP}
\end{equation}
\end{widetext}
The OMCP for $\ket\psi$ is just the pure state result as computed in section \ref{section:pure_states}.

\subsection{Isotropic states}
Isotropic states are those that are invariant under conjugation by any transformation of the form $U\otimes U^\ast$.\cite{Horodecki1999, Terhal2000a}  These are states of the form  
\begin{equation}
\rho = p\ket{\Phi^+}\bra{\Phi^+} + (1-p)\frac{\mathbbm{1}-\ket{\Phi^+}\bra{\Phi^+}}{d^2-1}\label{eq:def_iso_state}
\end{equation}
where $p\in[0,1]$ and $\ket{\Phi^+}$ is the maximally entangled state
\begin{equation}
\ket{\Phi^+} = \frac{1}{\sqrt d}\sum_{n=0}^{d-1} \ket{nn}.\label{eq:def_phi_p}
\end{equation}
The isotropic states are notable because they allow for substantial analytical progress in calculating the entanglement of formation for any $d$.\cite{Terhal2000}

For $p\geq 1/d^2$, this is in the form of a pure state plus colorless noise, as discussed in the previous section, with $x=(p-1/d^2)/(1-1/d^2)$.  The OMCP for a maximally entangled state is 1, so the OMCP for an isotropic state with $p\geq 1/d^2$ is $1/d + (p-1/d^2)(1-1/d)/(1-1/d^2)$.

The case of $p<1/d^2$ must be treated separately.  Consider the case of $x<0$ in equation \eqref{eq:pure_plus_noise_rho}.  When calculating the OMCP, the result is the same as in equation \eqref{eq:pure_plus_noise_OMCP} except that because $x$ is negative, when it is pulled out of the optimizations the minimization over $U_B$ becomes a maximization and the maximization over $U_A$ becomes a minimization.  The result, in the case that $\ket\psi=\ket{\Phi^+}$, is that the first term is exactly 0.

To see this, first note that an OMCP of 1 means that no matter what basis $B$ selects for his measurement, $A$ can always select a basis to guarantee that their measurements agree.  Instead of selecting this $U_A$, $A$ first applies this $U_A$ and then the permutation matrix sending $\ket0_A\mapsto\ket1_A\mapsto\cdots\mapsto\ket{d-1}_A\mapsto\ket0_A$, thus guaranteeing that her measurement will never agree with $B$'s.

Thus in the case of $p<1/d^2$, the result is just $(1-x)/d$ or $1/d + (1/d^2 - p)/(d-1/d)$.  Putting both cases together, the result is
\begin{equation}
\text{OMCP}=\frac{1}{d}+\frac{\left|p-\frac{1}{d^2}\right|}{1-1/d^2}\times
\left\{\begin{array}{cc}1/d, & \: p < 1/d^2\\ 1-1/d, & \: p \geq 1/d^2\end{array}\right.\,.\label{eq:OMCP_iso_evaluation}
\end{equation}

Note that these arguments apply equally well when $\ket{\Phi^+}$ is replaced by any other maximally entangled pure state, so the OMCP with such a replacement will be identical.  In particular, for $d=2$, replacing it by the singlet state gives the commonly studied class of Werner states, those which are invariant under conjugation by $U\otimes U$.  These states are separable for all $p<1/2$,\cite{Werner1989} demonstrating that the OMCP is in general not minimized on separable states.

\subsection{Two qubit states\label{section:two_qubits}}

Finally, I present exact results for all two qubit states.  As in the exact computation of the geometric discord,\cite{Dakic2010} the first step is to write the state $\rho$ in Bloch decomposed form:
\begin{equation}
\rho = \!\frac{1}{4}\!\!\left(\!\mathbbm{1}\!\otimes\!\mathbbm{1}\! +\!\! \sum_{i=1}^3 x_i \,\s_i\!\otimes\!\mathbbm{1}\! + \!\!\sum_{i=1}^3 y_i\, \mathbbm{1}\!\otimes\!\s_i\! + \!\!\sum_{i,j=1}^3 t_{ij}\,\s_i\!\otimes\!\s_j\!\right)\label{eq:Bloch_decomp}
\end{equation} 
where $\mathbbm{1}$ is the $2\times 2$ identity matrix, the $\s_i$ are the three Pauli matrices, $x_i=\text{Tr}\big(\rho\,(\s_i\!\otimes\!\mathbbm{1})\big)$, $y_i=\text{Tr}\big(\rho\,(\mathbbm{1}\!\otimes\!\s_i)\big)$, and $t_{ij}=\text{Tr}\big(\rho\,(\s_i\otimes\s_j)\big)$.  For a density matrix of the form $\rho=\rho_A\otimes\rho_B$, the MCP for a specified $U_A$ and $U_B$ is $\text{Tr}\left(U_A^{ }\rho_A U_A^\dg \,\circ \,U_B^{ }\rho_B U_B^\dg\right)$; the proof is similar to the one given in appendix \ref{appendix:OMCP_simple_proof} for equation \eqref{eq:OMCP_simple_2}.  This can be applied individually to each term in equation \eqref{eq:Bloch_decomp}.  The first term is invariant under any unitary conjugation, giving $\text{Tr}(\mathbbm{1}\circ\mathbbm{1})=2$.  The second and third terms both give 0, since for any matrix $M$, $\text{Tr}(\mathbbm{1}\!\circ\!M) = \text{Tr}(M\!\circ\!\mathbbm{1}) = \text{Tr}(M)$, and $\text{Tr}(U\s_i U^\dg) = \text{Tr}(\s_i)=0$.  Thus 
\begin{equation}
\text{MCP}=\frac{1}{4}\left[2 + \sum_{ij} t_{ij}\text{Tr}\left(U_A^{ }\s_i\,U_A^\dg\circ U_B^{ }\s_j\,U_B^\dg\right)\right].
\end{equation}
This can now be explicitly computed for fully general unitaries of the form
\begin{equation}
U = \left(\begin{array}{cc}
\cos(\t)e^{i\phi} & \sin(\t)e^{i\psi}\\ 
-\sin(\t)e^{i(\varphi-\psi)} & \cos(\t)e^{i(\varphi-\phi)}
\end{array}\right).
\end{equation}
Carrying out this computation gives
\begin{equation}
\text{MCP}=\frac{1}{2}\left[1 + \hat{r}_A^T\cdot T\cdot\hat{r}_B^{ }\right]
\end{equation}
where $T$ is the matrix with entries $t_{ij}$ and 
\begin{align}
\hat{r}_A^{ } &= \left(\begin{array}{c}\sin(2\t_A)\cos(\phi_A-\psi_A)\\ \sin(2\t_A)\sin(\phi_A-\psi_A) \\ \cos(2\t_A)\end{array}\right),\\
\hat{r}_B^{ } &= \left(\begin{array}{c}\sin(2\t_B)\cos(\phi_B-\psi_B)\\ \sin(2\t_B)\sin(\phi_B-\psi_B)\\ \cos(2\t_B)\end{array}\right).
\end{align}
The two unit vectors $\hat{r}_A^{ }$ and $\hat{r}_B^{ }$ are determined by the unitaries $U_A$ and $U_B$ respectively, and furthermore this map from unitaries to unit vectors is surjective, so the optimizations over $U_A$ and $U_B$ can be replaced with unconstrained optimizations over $\hat{r}_A$ and $\hat{r}_B$:
\begin{equation}
\text{OMCP}=\frac{1}{2}\left[1+\min_{\hat{r}_B}\left(\max_{\hat{r}_A}\left(\hat{r}_A^T\cdot T\cdot\hat{r}_B^{ }\right)\right)\right].
\end{equation}
Next, the matrix $T$ is real and hence has a singular value decomposition $T=O_1D\,O_2^T$ where $O_1$ and $O_2$ are real orthogonal matrices and $D$ is diagonal, with its diagonal elements being the singular values of $T$.  But multiplying an orthogonal matrix by a unit vector gives some new unit vector, and since the unit vectors are both optimized over the full unit sphere, this means we can assume without loss of generality that $O_1=O_2=\mathbbm{1}$.

The optimization can now be carried out explicitly.  For any $\hat{r}_B$, $A$ can choose $\hat{r}_A=\hat{r}_B\equiv\hat{r}$, in which case $\hat{r}^T\cdot T \cdot \hat{r}$ becomes $\sum_i \hat{r}_i^2 d_i$, where $d_1,\,d_2,\,d_3\geq 0$ are the singular values of $T$. Because $\hat{r}$ is a unit vector, this is a weighted sum of the three singular values, which $B$ can minimize by putting all the weight on the smallest one.  In other words, 
\begin{equation}
\min_{\hat{r}_B}\left(\!\max_{\hat{r}_A}\left(\hat{r}_A^T\cdot T\cdot\hat{r}_B^{ }\right)\!\right) \geq \min_{\hat{r}}\left(\hat{r}^T\cdot T\cdot\hat{r}\right) = \min(\{d_i\}).\label{eq:two_qubit_geq}
\end{equation}
On the other hand, suppose that $B$ chooses $\hat{r}_B$ to lie in the direction corresponding to the smallest singular value; eg. if $d_1\leq d_2,d_3$, then $\hat{r}_B=(1,0,0)^T$.  In that case, $\hat{r}_A^T\cdot T\cdot\hat{r}_B^{ }$ becomes $\hat{r}^A_1 d_1$ which $A$ maximizes by choosing $\hat{r}_A=(1,0,0)^T$ as well, giving $d_1$.  Thus
\begin{align}
\min_{\hat{r}_B}\left(\!\max_{\hat{r}_A}\left(\hat{r}_A^T\cdot T\cdot\hat{r}_B^{ }\right)\!\right) & \leq \left(\!\max_{\hat{r}}\left(\hat{r}_j\,d_j\right) \middle| d_j = \text{min}(\{d_i\})\!\right)\nonumber\\& =\text{min}(\{d_i\}).\label{eq:two_qubit_leq}
\end{align}
The two inequalities \eqref{eq:two_qubit_geq} and \eqref{eq:two_qubit_leq} together prove that the optimization over $\hat{r}_A$ and $\hat{r}_B$ gives exactly the minimum singular value of the matrix $T$.

I now summarize the result: given any $4\times 4$ density matrix $\rho$, the OMCP is found by the following steps:
\begin{enumerate}
\item Compute the matrix $T$ with elements $t_{ij} = \text{Tr}\big(\rho\,(\s_i\otimes\s_j)\big)$
\item Find the smallest singular value of $T$; call it $s$
\item $\text{OMCP}=(1+s)/2$; $\text{accord}=s$
\end{enumerate}
Although it is not strictly speaking an analytical expression, this is an extremely efficient numerical computation.

One important consequence of this result is that the OMCP and accord are symmetric in the two subsystems for $d=2$: when $A$ and $B$ are swapped, the result is to send $T$ to $T^T$, but any matrix and its transpose have the same singular values.

One might hope to extend this approach to higher dimensions.  The three Pauli matrices in equation \eqref{eq:Bloch_decomp} can be replaced with corresponding higher-dimensional traceless Hermitian matrices, such as the Gell-Mann matrices for $d=3$,\cite{GellMann1962} and one can generate the $(d^2-1)\times(d^2-1)$ analogue of $T$; the natural conjecture is that the accord is again just the smallest singular value of this matrix.  Although this is true in certain special cases, such as the isotropic states discussed above for which all the singular values are equal, it is sadly not true in general.  The reason is that, although for any $d$ the optimization over unitaries can be rewritten in terms of an optimization over $(d^2-1)$-dimensional unit vectors, in general this mapping from the unitary group $\text{U}(d)$ to the orthogonal group $\text{O}(d^2-1)$ is not surjective; consequently, the optimization over the unit vectors depends on the allowed regions of the hypersphere and on how those regions are transformed by the matrices $O_1$ and $O_2$ from the singular value decomposition of $T$.  Thus the simple reasoning used for $d=2$ above no longer applies.


\section{Comparison with other measures\label{section:comparison}}

The accord can be compared with commonly used measures of entanglement and of quantum correlations more generally.  In particular, I consider the four notions of quantum correlations discussed in the introduction: nonlocality, as seen in violations of Bell-type inequalities; entanglement, as captured by the entanglement of formation; teleportation fidelity; and discord.  These measures and the OMCP all agree on pure states, in the sense that each one is a function only of the Schmidt coefficients of the state\cite{Vidal2000} and furthermore that each is minimized on the same set of states, namely product states.  Notably, the teleportation fidelity is linearly related to the singlet fraction\cite{Horodecki1999b}, which for pure states is identical to the OMCP, $\left(\sum c_i\right)^2/d$,\cite{Banaszek2000}  and the concurrence\cite{Hill1997,Wootters1998} for two qubit states, which is in bijection with the entanglement of formation, is identical to the accord. (See Appendix \ref{appendix:sing_frac} for a simple proof of the pure state singlet fraction.)  Note that one additional common measure, the 1/2-R\'{e}nyi entropy\cite{Renyi1961}, evaluates to $2\log(\sum c_i)$ and is thus also closely related to the accord.

\begin{figure*}
\centering
\includegraphics[width=\textwidth]{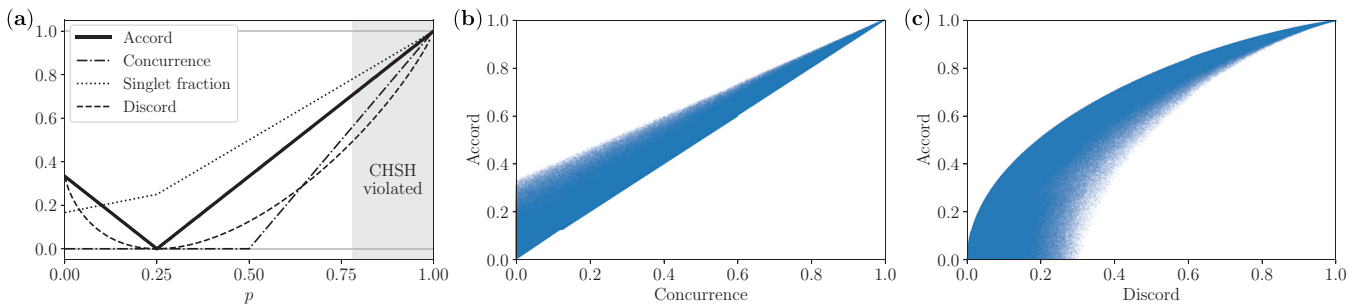}
\caption{{\bf(a)} Various measures of quantum correlations computed for isotropic states of the form given in equation \eqref{eq:def_iso_state} with $d=2$.  Entanglement, teleportation fidelity, nonlocality, and nonclassicality are captured by concurrence, singlet fraction, violations of the CHSH inequality, and the discord respectively.  The accord appears similar to the discord.  {\bf(b)} Accord ($A$) versus concurrence ($C$) for $10^6$ randomly generated Bell diagonal states.  Evidently these satisfy $C(\rho)\leq A(\rho)$.  {\bf(c)} Accord versus discord ($D$) for $10^6$ randomly generated Bell diagonal states.  These satisfy $J\big(A(\rho)\big)\leq D(\rho)$ where the function $J$ is defined in equation \eqref{eq:J}.\label{fig:measure_comparison}}
\end{figure*}

For mixed states, the various measures are no longer equivalent.  To illustrate this fact, and to situate the accord among the existing measures, I evaluate each one exactly for the isotropic states as defined in equation \eqref{eq:def_iso_state} with $d=2$; the results are shown in Figure \ref{fig:measure_comparison}(a).  Nonlocality is demonstrated by the violation of the CHSH inequality for $p > (1+3/\sqrt{2})/4$\cite{Horodecki1995,Horodecki2009}.\footnote{There are other more complicated inequalities which show that the actual boundary between states that do and do not admit a LHVM is at a lower value of $p$\textcolor{blue}{$^{66}$}, though not lower than 0.74\textcolor{blue}{$^{67}$}}$^,$\cite{Vertesi2008,Acin2006}  Entanglement is shown via the concurrence, which can be computed efficiently.\cite{Hill1997,Wootters1998}  Teleportation fidelity is replaced by the linearly related singlet fraction to emphasize that the latter is no longer equal to the OMCP on mixed states.  The singlet fraction and the discord are computed exactly using results from references \onlinecite{Badziag2000},\onlinecite{Verstraete2002} and \onlinecite{Luo2008},\onlinecite{Modi2012} respectively.  

On the isotropic states, accord and discord appear quite similar: they are both zero only for the maximally mixed state, and they even share their maximum values at $p=0$ and $p=1$.  In fact, the two are related by a simple formula for the $d=2$ isotropic states; the discord is $D=I-J(a)$ where $I$ is the quantum mutual information and 
\begin{equation}
J(a)=\big[(1+a)\log_2(1+a) + (1-a)\log_2(1-a)\big]/2\label{eq:J}
\end{equation} 
where $a$ is the accord.\cite{Luo2008,Modi2012}  

This does not generalize.  In fact, there are other classes of states where the accord appears to match the entanglement instead.  To demonstrate this, I consider the class of states $\rho$ that are diagonal in the Bell basis:
\begin{subequations}
\begin{align}
\ket{\Phi^\pm} & = \frac{1}{\sqrt 2}\left(\ket{00}\pm\ket{11}\right)\\
\ket{\Psi^\pm} & = \frac{1}{\sqrt 2}\left(\ket{01}\pm\ket{10}\right)
\end{align}
\end{subequations}
In particular, consider Bell diagonal states with diagonal elements $(1/2,x/2,(1-x)/2,0)$.  For all states of this type, both concurrence (and therefore any entanglement measure) and accord are exactly zero, while the discord is
\begin{equation}
\frac{3}{2} + \frac{x}{2}\log_2\left(\frac{x}{2}\right) + \frac{1-x}{2}\log_2\left(\frac{1-x}{2}\right) - J\left(\frac{1}{2}+\left|x-\frac{1}{2}\right|\right)
\end{equation}
for $J$ defined above.  

\begin{figure*}
\centering
\includegraphics[width=\textwidth]{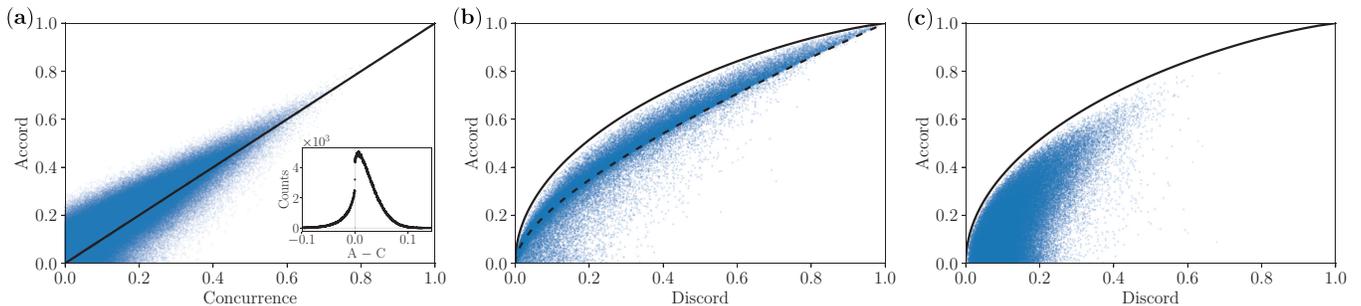}
\caption{ {\bf(a)} Accord versus concurrence for $10^6$ arbitrary two qubit states found by tracing out two sites of random four qubit pure states\cite{Randstates1}; a line showing $C(\rho)= A(\rho)$ is provided as a guide to the eye.  The inset shows the distribution of $(\text{accord}-\text{concurrence})$, computed as a histogram with $10^3$ bins.   Surprisingly, it is possible for states to be entangled while having no measurement correlations. {\bf(b)} Accord versus discord, computed by numerical optimization, for $10^5$ arbitrary two qubit states, again found by tracing out sites from four qubit pure states.\cite{Randstates2}  The solid line shows the conjectured bound $J\left(A(\rho)\right)\leq D(\rho)$; the bound is never violated. The dashed line shows the relation for pure states.  {\bf(c)} Accord versus discord for $10^5$ two qubit states drawn from a different distribution.\cite{Randstates1} \label{fig:ent_comparison_all}}
\end{figure*}

The two classes of states together suggest that $\{\rho | C(\rho)=0\}\supset\{\rho | A(\rho)=0\}\supset\{\rho | D(\rho)=0\}$, where $C$, $A$, and $D$ are the concurrence, accord, and discord.  Informally, this could be summarized by saying that the accord is an intermediate measure between entanglement and discord.  As a further demonstration, I show in Figure \ref{fig:measure_comparison}(b) and (c) the accord versus the concurrence and versus the discord for $10^6$ randomly generated Bell diagonal states; for all such states, $C(\rho)\leq A(\rho)$ and $J(A(\rho))\leq D(\rho)$ where $J$ is given by equation \eqref{eq:J}.  In this sense the accord is both an upper bound on the entanglement and a lower bound on the discord, at least for this class of states.

The story is less clear when considering all two qubit states, for which the inequality $C(\rho)\leq A(\rho)$ is violated.  This is demonstrated in Figure \ref{fig:ent_comparison_all}(a), which shows concurrence versus accord for $10^6$ states found by tracing out two sites in randomly generated four qubit pure states.\cite{Randstates1}  There was no particular reason why that inequality should hold in general, so its violation is not in itself surprising.  On the other hand, the figure also reveals that there are states with zero accord and nonzero entanglement, which is quite unexpected.  Recalling the definition of the accord, this means that it is possible to choose a local basis for one subsystem that completely negates the correlations between measurements made on the two qubits.  This seems to suggest that the entanglement in such states is somehow hidden or less useful; although reminiscent of the idea of bound entanglement that cannot be used for distillation\cite{Horodecki1998}, it must not be equivalent because all entangled states of two qubits are distillable\cite{Horodecki1997}.


This surprising result can be clarified somewhat by further examination of the Bloch representation, equation \eqref{eq:Bloch_decomp}.  The accord does not depend at all on the coefficients $x_i$ or $y_i$; in fact, these coefficients precisely determine the reduced density matrices on the two subsystems: $\rho_A^{ } = \left(\mathbbm{1} + \mathbf{x}\cdot \bm{\s}\right)/2$ and $\rho_B^{ } = \left(\mathbbm{1} + \mathbf{y}\cdot \bm{\s}\right)/2$.  Thus the accord has no dependence whatsoever on the local states, only on the correlations; this applies not just to two qubit states, but to any $d$.  The same is not true for the entanglement (nor for the discord), so it seems likely that entanglement in states with no unavoidable measurement correlations in fact comes from the coefficients $\{x_i\}$ and $\{y_i\}$, ie. from the local reduced density matrices.  This is indeed the case.  Any state with those coefficients equal to 0 is equivalent under local unitaries to a Bell diagonal state,
\footnote{To see this, first perform the singular value decomposition of the matrix $T$ as in section \ref{section:two_qubits}; the matrices $O_1$ and $O_2$ correspond to local unitaries, so we can assume that $T$ is in fact diagonal.  This corresponds to a Bell diagonal state with diagonal elements $(1+t_{11}+t_{22}-t_{33})/4$, $(1+t_{11}-t_{22}+t_{33})/4$, $(1-t_{11}+t_{22}+t_{33})/4$, and $(1-t_{11}-t_{22}-t_{33})/4$.} 
and Figure \ref{fig:measure_comparison}(b) clearly demonstrates that for any such state $C(\rho) \leq A(\rho)$.


The lower bound on the discord, on the other hand, appears to hold even for arbitrary states, though the evidence is only numerical.  I again generate random two qubit states by tracing out two qubits from four qubit pure states.  Drawing $10^5$ random states from each of five different distributions for the four qubit states, and in each case computing the discord of each mixed state by numerical optimization, the bound $J\left(A(\rho)\right) \leq D(\rho)$ is never violated.  I show accord versus discord for two of the pure state distributions\cite{Randstates1,Randstates2} in Figure \ref{fig:ent_comparison_all}(b) and (c), with a line giving the bound for comparison.


\section{Discussion\label{section:discussion}}

In this work, I have introduced a new measure of quantum correlations, the accord, defined by a simple thought experiment.  I have computed its value for a pure state of two $d$-dimensional subsystems to be $\left[\left(\sum c_i\right)^2-1\right]\big/(d-1)$, where the $c_i$ are the Schmidt coefficients of the state, and I have furthermore explicitly computed the value on several important classes of mixed states, including all states of two qubits and all pure states plus colorless noise.

For pure states the accord is closely related to the maximal singlet fraction and hence to the teleportation fidelity, and it is also a simple function of the 1/2-R\'{e}nyi entropy.  For two qubit pure states it is equal to the concurrence.  

For mixed states the accord lies approximately between entanglement and discord. In particular, there are classes of states for which entanglement is zero while both accord and discord are generically nonzero, and also classes for which entanglement and accord are both zero while discord is not.  In fact, for all two qubit Bell diagonal states, or equivalently all states with a maximally mixed local density matrix for each of the two qubits, zero accord implies zero entanglement, and zero discord implies zero accord; on these states, accord is a tight upper bound on the concurrence and is in bijection with a tight lower bound on the discord.

The statement that accord lies between entanglement and discord becomes approximate when considering the more general case of all mixed states.  For arbitrary two qubit states, the accord continues to provide a tight lower bound on discord, but the bound on entanglement is sometimes violated.  Most notably, there are states with nonzero entanglement and zero accord.  In other words, there exist entangled states of two qubits for which, when one qubit is measured in a particular basis, measurements of the two qubits will be completely uncorrelated regardless of the measurement basis chosen for the second qubit.  This remarkable and unexpected behavior reveals a new type of hidden entanglement that is possible only in mixed states. 

This novel insight into the nature of mixed state entanglement demonstrates the value of studying measures like the accord that are designed to be as intuitively clear as possible.  In particular, the accord formalizes one of the most intuitive pictures of entanglement, based on the correlations between measurements of two subsystems, and hence a comparison between the accord and other measures like entanglement and discord allows a clear picture of how they do or do not conform to this kind of intuitive understanding.  This, combined with the fact that it can be efficiently computed on several very important classes of states, makes the accord a valuable and interesting measure of quantum correlations.  

One additional benefit arising from its definition in terms of a clear experimental procedure is that the accord is relatively easy to explain to beginning students of quantum mechanics and even to non-physicists, certainly compared with most measures of entanglement, and thus it could also prove useful for education and outreach purposes.  

There is certainly more work remaining to be done.  It may be possible to make progress on the efficient evaluation of the accord for general mixed states beyond the two qubit case by a careful consideration of the hypersphere geometry alluded to at the end of section 
\ref{section:two_qubits}.  Additionally, in this paper I have computed the accord only for the case where both subsystems $\mathcal{H}_A$ and $\mathcal{H}_B$ have the same dimension, $d$; it would be interesting to pursue the more general case of $d_A\neq d_B$ as mentioned in section \ref{section:unequal_dims}.  


\begin{acknowledgments}
I would like to thank Quntao Zhuang and Yichen Huang for helpful comments on the manuscript, and Eric Dodds and Lena Evans for pointing out the existence of literature on Hadamard products.  I would also like to thank Andrew Szasz for asking me to explain, in a way understandable to a non-physicist, what it means for one entangled state to be more or less entangled than another; my attempts to answer this question led directly to the definition of the accord.  I am supported by the U.S. Department of Energy, Office of Science, Office of Basic Energy Sciences, Materials Sciences and Engineering Division under Contract No.\ DE-AC02-05-CH11231 through the Scientific Discovery through Advanced Computing (SciDAC) program (KC23DAC Topological and Correlated Matter via Tensor Networks and Quantum Monte Carlo).
\end{acknowledgments}




\appendix
\section{Pure state OMCP simplification\label{appendix:OMCP_simple_proof}}

Here I show that for a pure state $\rho = \ket\psi\bra\psi$, where $\ket\psi$ has Schmidt coefficients $\{c_i\}$, equation \eqref{eq:def_OMCP} is equivalent to the simpler formulations given in equations \eqref{eq:OMCP_simple_1} and \eqref{eq:OMCP_simple_2}.  

\begin{figure}
\centering
\includegraphics[width=0.48\textwidth]{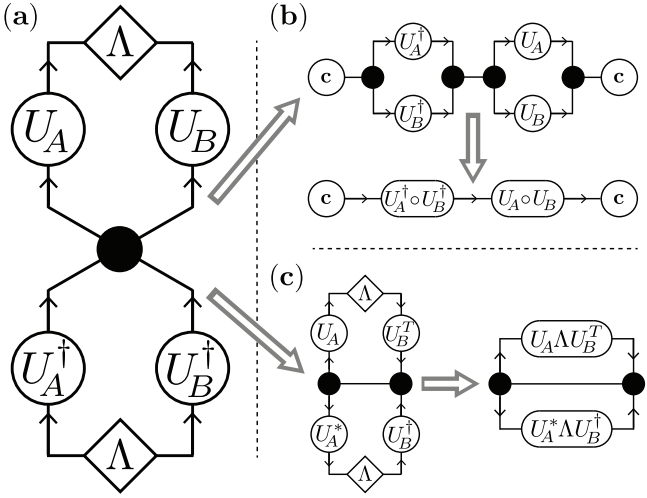}
\caption{{\bf(a)} Tensor network representation of equation \eqref{eq:pure_MCP}.  Each node is a tensor, and lines indicate tensor contraction.  Filled circles are higher dimension identity tensors, the diamonds are diagonal matrices whose diagonal entries are the Schmidt coefficients, and for nondiagonal matrices inward arrows indicate the row index and outward arrows the column index. {\bf(b)} Derivation of equation \eqref{eq:OMCP_simple_1} using the identities from Figure \ref{fig:TN_identities}.  The circle labeled {\bf{c}} is a vector whose entries are the Schmidt coefficients.  {\bf(c)} Derivation of equation \eqref{eq:OMCP_simple_2}.\label{fig:OMCP_pure}}
\end{figure}

For this special case, the MCP becomes
\begin{equation}
\sum_{ijn}c_ic_j\bra{i,i}(U_A\otimes U_B)^\dg\ket{n,n}\bra{n,n}(U_A\otimes U_B)\ket{j,j}.\label{eq:pure_MCP}
\end{equation}
This expression can be written as the tensor network shown in Figure \ref{fig:OMCP_pure}(a); edges in the network indicate tensor contraction, and for nondiagonal matrices the arrows point inwards for the row index and outwards for the column index.  Each diamond tensor $\Lambda$ is the diagonal matrix with diagonal entries $c_1$ through $c_d$, and the filled circle represents a higher dimensional identity tensor, for which any element with all indices equal is 1 and all others are 0.

\begin{figure}
\centering
\includegraphics[width=0.48\textwidth]{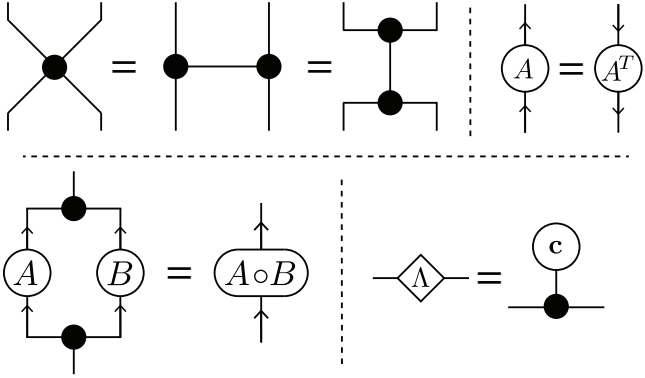}
\caption{Tensor network identities used in deriving equations \eqref{eq:OMCP_simple_1} and \eqref{eq:OMCP_simple_2}.  The notation used is defined in the caption of Figure \ref{fig:OMCP_pure}.\label{fig:TN_identities}}
\end{figure}

This tensor network can now be manipulated using the identities shown in Figure \ref{fig:TN_identities}.  Following the steps shown in Figure \ref{fig:OMCP_pure}(b) immediately gives equation \eqref{eq:OMCP_simple_1}, and similarly following the steps in Figure \ref{fig:OMCP_pure}(c) gives equation \eqref{eq:OMCP_simple_2}.


\begin{widetext}
\section{Proof of Hadamard trace lemma\label{appendix:Had_tr_pf}}

Here I present an alternative proof of equation \eqref{eq:Had_trace_lemma} to the one given in reference \onlinecite{Taskara2013}.  The proof proceeds by direct expansion of the second term on the right-hand side:
\begin{subequations}
\begin{align}
\sum_{i=1}^{d-1}\sum_{j=i+1}^d(a_{ii}-a_{jj})(b_{ii}-b_{jj}) & = \sum_{i=1}^{d-1}\sum_{j=i+1}^d a_{ii}b_{ii} + \sum_{i=1}^{d-1}\sum_{j=i+1}^d a_{jj}b_{jj} - \sum_{i=1}^{d-1}\sum_{j=i+1}^d a_{ii}b_{jj} - \sum_{i=1}^{d-1}\sum_{j=i+1}^d a_{jj}b_{ii}\\
& = \sum_{i=1}^{d-1}\sum_{j=i+1}^d a_{ii}b_{ii} + \sum_{j=2}^{d}\sum_{i=1}^{j-1} a_{jj}b_{jj} - \sum_{i=1}^{d-1}\sum_{j=i+1}^d a_{ii}b_{jj} - \sum_{j=2}^{d}\sum_{i=1}^{j-1} a_{jj}b_{ii}\\
& = \sum_{i=1}^{d-1}(d-i) a_{ii}b_{ii} + \sum_{i=2}^{d}(i-1) a_{ii}b_{ii} - \sum_{i=1}^{d-1}\sum_{j=i+1}^d a_{ii}b_{jj} - \sum_{i=2}^{d}\sum_{j=1}^{i-1} a_{ii}b_{jj}\\
& = (d-1)\sum_{i=1}^d a_{ii}b_{ii}- \sum_{i=1}^{d}\sum_{j\neq i} a_{ii}b_{jj} \\
& = d\,\sum_{i=1}^d a_{ii}b_{ii}- \sum_{i=1}^{d}\sum_{j=1}^d a_{ii}b_{jj}\\
& = d\,\text{Tr}(A\circ B) - \text{Tr}(A)\,\text{Tr}(B)
\end{align}
\end{subequations}
This is precisely the desired result.
\end{widetext}

\begin{figure}
\centering
\includegraphics[width=0.48\textwidth]{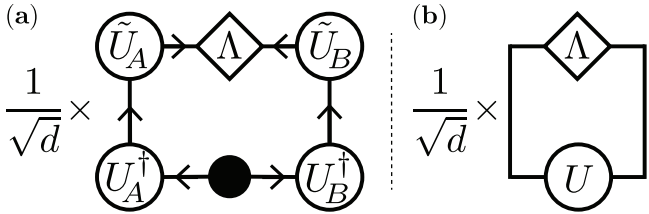}
\caption{{\bf(a)} Tensor network representation of $\left\langle \psi_{\text{m}} | \psi \right\rangle$ from the calculation of the singlet fraction.  {\bf(b)} The network immediately simplifies to this one, where $U$ is some unitary matrix.\label{fig:singlet_fraction}}
\end{figure}

\section{Pure state singlet fraction\label{appendix:sing_frac}}

I include this calculation of the singlet fraction for pure states primarily to demonstrate the power of the tensor matrix notation used in Appendix \ref{appendix:OMCP_simple_proof}, which allows for an almost trivial proof.

For a pure state $\ket\psi$, the singlet fraction is the maximum over all maximally entangled states $\ket{\psi_\text{m}}$ of $\left|\left\langle \psi_{\text{m}} | \psi \right\rangle\right|^2$.  There exist $\tilde{U}_A$, $\tilde{U}_B$, $U_A$, and $U_B$ such that $\ket\psi$ satisfies equation \eqref{eq:def_Schmidt_decomp} and $\ket{\psi_\text{m}} = (U_A\otimes U_B)\ket{\Phi^+}$, where $\ket{\Phi^+}$ is defined in equation \eqref{eq:def_phi_p}.  The maximization over $\ket{\psi_\text{m}}$ becomes a maximization over $U_A$ and $U_B$.

Then the inner product $\left\langle \psi_{\text{m}} | \psi \right\rangle$ is given by the tensor network shown in Figure \ref{fig:singlet_fraction}(a).  The lower filled circle is just the identity matrix and can be removed from the network.  Then all four unitary matrices can be multiplied together, giving some new unitary $U$; the inner product is now given by the network in \ref{fig:singlet_fraction}(b), which is just $\text{Tr}(U\Lambda)/\sqrt{d} = (\sum_i U_{ii}c_i)/\sqrt{d}$.  But 
\begin{equation}
\left|\sum_i U_{ii}c_i\right| \leq \sum_i\left|U_{ii}c_i\right| \leq \sum_i c_i
\end{equation}
and both bounds are achieved for $U=\mathbbm{1}$.  Thus the maximum of $\left|\left\langle \psi_{\text{m}} | \psi \right\rangle\right|^2$ is $(\sum c_i)^2/d$, which as claimed in the main text exactly matches the result for the OMCP.


\section{Entangled state with no accord\label{appendix:state}}
In section \ref{section:comparison} above, I showed the surprising fact that states can have quite large entanglement even with nearly zero accord.  In fact, it is possible to find a state with precisely zero accord and nonzero entanglement.  To generate such a state, I begin with a nonnegative $3\times3$ diagonal matrix $T$ with one of the diagonal entries being 0 and the other two random.  I then compute $O_1 T\,O_2$ for random orthogonal matrices $O_1$ and $O_2$, and also randomly pick values for $\{x_i\}$ and $\{y_i\}$.  Then the state $\rho$ using equation \eqref{eq:Bloch_decomp} is guaranteed to have zero accord.  After checking that this is a valid density matrix, ie. it is positive semidefinite, the concurrence can be computed.  One example state $\rho$ found by this method is given by:
\begin{widetext}
\begin{equation}
\rho = \left(\begin{array}{cccc}
0.1547077 & -0.0937756-0.0097791i & 0.0032410-0.0780971i & -0.0490784-0.0004913j\\
-0.0937756+0.0097791i & 0.2401018 & 0.1384087 & 0.0790484-0.0248949i\\
0.0032410+0.0780971i & 0.1384087 & 0.1802319 & -0.0179682+0.0434231i\\
-0.0490784+0.0004913i & 0.0790484+0.0248949i & -0.0179682-0.0434231i & 0.4249586
\end{array}\right).
\end{equation}
The interested reader can check that this state indeed has the claimed properties.  (Note that due to rounding, the accord is about $3\times 10^{-8}$ rather than exactly 0.) 
\end{widetext}


\bibliography{measurement_entanglement_bib}

\begin{thebibliography}{75}
\bibitem [{Ran({\natexlab{a}})}]{Randstates1}%
  \BibitemOpen
  \bibinfo {note} {The pure states are generated by picking an amplitude for
  each of the 16 coefficients from a uniform distribution from 0 to 1 and
  likewise for the phase from 0 to $2\pi$, then normalizing at the
  end.}\BibitemShut {Stop}%
\bibitem [{Ran({\natexlab{b}})}]{Randstates2}%
  \BibitemOpen
  \bibinfo {note} {The 16 coefficient amplitudes are drawn from an
  exponentially decaying distribution, $\left\langle
  \left|a_i\right|^2\right\rangle\approx 2^{-i}$. Phases are uniformly
  distributed for each coefficient and the state is normalized at the
  end.}\BibitemShut {Stop}%
\end{thebibliography}


\end{document}